\documentclass[aps,twocolumn,prl,showpacs,floatfix,superscriptaddress]{revtex4}

\usepackage{amsmath,fixmath}
\usepackage{graphicx}
\usepackage{acronym}
\usepackage{amssymb}
\usepackage{bm}

\begin{document}

\title{Phase diagram of a polydisperse soft-spheres model for liquids
  and colloids } 

\author{L.~A.~Fern\'andez} \affiliation{Departamento
  de F\'\i{}sica Te\'orica I, Universidad
  Complutense,
  28040 Madrid, Spain.}  \affiliation{Instituto de Biocomputaci\'on y
  F\'{\i}sica de Sistemas Complejos (BIFI), Spain.}

\author{V.~Mart\'\i{}n-Mayor} \affiliation{Departamento de F\'\i{}sica
  Te\'orica I, Universidad Complutense,
  28040 Madrid, Spain.}  \affiliation{Instituto de Biocomputaci\'on y
  F\'{\i}sica de Sistemas Complejos (BIFI), Spain.}

\author{P.~Verrocchio} \affiliation{Instituto de Biocomputaci\'on y
  F\'{\i}sica de Sistemas Complejos (BIFI), Spain.}  \date{\today}
\affiliation{Dipartimento di Fisica, Universit\`a di Trento and CRS,
  SOFT, INFM-CNR, 38050 Povo, Trento, Italy.}

\begin{abstract}
  The phase diagram of soft spheres with size dispersion has been
  studied by means of an optimized Monte Carlo algorithm which allows
  to equilibrate below the kinetic glass transition for all sizes
  distribution. The system ubiquitously undergoes a first order
  freezing transition. While for small size dispersion the frozen
  phase has a crystalline structure, large density inhomogeneities
  appear in the highly disperse systems.  Studying the interplay
  between the equilibrium phase diagram and the kinetic glass
  transition, we argue that the experimentally found terminal
  polydispersity of colloids is a purely kinetic phenomenon.
\end{abstract}
\pacs{64.60.Fr,64.60.My,66.20.+d} 
\maketitle 

The equilibrium phase diagram of dense classical fluids is far from
being fully understood, especially as regards the influence over the
freezing transition of some disorder in the parameters of the
interaction. While fluids made of identical atoms crystallize easily
upon lowering the temperature or increasing the density, the fate of
{\em polydisperse} systems (colloids, for instance), where the
particle size $\sigma$ is sampled from a probability distribution,
$P(\sigma)$, is still a matter of debate.

On the theoretical side, the effect of size dispersion, measured by
the quantity $\delta$ (the ratio among the standard deviation and the
mean of $P(\sigma)$) over the phase diagram has been addressed mostly
in the case of the hard-spheres model, which is meant to describe
colloidal systems~\cite{Sciortino05}. At least for small
polydispersity, $\delta$ seems to be the only feature of $P(\sigma)$
that controls the physical results. Different theories predict
conflicting scenarios in the region of large $\delta$.  The moment
free-energy method~\cite{Fasolo04} suggests phase separation between
many different crystal phases, each one with a much narrower size
dispersion than $\delta$ (a phenomenon termed {\em
fractionation}). However, a density functional
analysis~\cite{Chaudhuri05} predicts the existence of a terminal
polydispersity $\delta_\mathrm{t}$ beyond which the formation of the
crystal is thermodynamically suppressed.  Numeric simulations of the
hard sphere model find some agreement with both the
fractionation~\cite{Bartlett98,Kofke99} and the terminal
polydispersity~\cite{Auer01} scenarios. As regards models with a soft
potential (e.g. Lennard-Jones), which are more appropriate to describe
liquids, no extensive study on the effects of polydispersity has been
performed so far.

On the experimental side, crystallization of very viscous colloidal
samples with $\delta > \delta_\mathrm{t}\approx 0.12$ does not occur,
even after months spent from the preparation~\cite{Pusey86}. Further
evidence supporting the terminal polydispersity scenario comes from
the finding of reentrant melting (crystal to fluid transition when
increasing the density) on polydisperse colloids in confined
geometry~\cite{Dullens04}.

Yet, these experimental results do not necessarily reveal features of
the phase diagram. Indeed, the processes needed to keep the system in
thermodynamic equilibrium often become exceedingly slow.  Such
processes are the nucleation of the solid within the fluid and the
$\alpha$-relaxation in the super-cooled fluid~\cite{DeBenedetti97}.

Here we obtain the {\em equilibrium} phase diagram of polydisperse
soft-spheres in the $(N,V,T)$ ensemble, aiming to describe both
liquids and colloids.  This is made possible by the combination of an
optimized Monte Carlo (MC) method (which, unlike standard MC,
equilibrates {\em below} the kinetic glass temperature) and a novel
Finite-Size Scaling study of the fluid-solid coexistence line.  At all
$\delta$, a first-order freezing transition separates the fluid phase
from the low temperature solid. This rules out the thermodynamic
origin of the terminal polydispersity scenario.  However, we show that
a Brownian dynamics will not find crystallization for $\delta>0.12$,
in quantitative agreement with colloids
experiments~\cite{Pusey86}. Furthermore, depending on polydispersity
the solid phase can be either a crystal or an inhomogeneous phase
(hereafter I-phase). The freezing temperature shows a reentrant
behavior when increasing $\delta$. Interestingly, in the range
$\delta\in[0.12,0.38]$ the kinetic glass transition occurs for
temperatures, $T$, and densities, $\rho$, in the stable fluid phase.

Specifically, in our simulations soft-spheres of radius $\sigma_i$ and
$\sigma_j$ at distance $r$ interact via the pair 
potential~\footnote{We use the long distance cut-off
of~\cite{Fernandez06c,yan04}. Our length unit $\sigma_0$ is fixed by
$\sigma_0^3 \!=\! \int \mathrm{d}\sigma_i \mathrm{d}\sigma_j
P(\sigma_i) P(\sigma_j) (\sigma_i + \sigma_j)^3$. We simulated $N$
(initially fully disordered) particles in a box with periodic boundary
conditions at $\rho\!=\!\sigma_0^{-3}$.}
\begin{equation}
V_{ij}(r) = \left(\frac{\sigma_i+\sigma_j}{r} \right)^{12}\,.
\label{potential}
\end{equation}
The effect of $T$ and $\rho$ is encoded in the single thermodynamic
parameter $\Gamma \equiv \rho T^{-1/4}$.  Although ~(\ref{potential})
generalizes well known models for simple liquids~\cite{hansen}, its
scale-invariant form suggests that it could describe as well colloids,
whose size is in the micrometer range. We consider a flat size
distribution, constant in the range
$[\sigma_\mathrm{min},\sigma_\mathrm{max}]$.  In order to eliminate
sample-to-sample fluctuations we follow~\cite{Santen01}, which for a
flat distribution amounts to pick $\sigma_i \!=\! \sigma_\mathrm{min}
+ \Delta(i-1)$, with
$\Delta\!=\!(\sigma_\mathrm{max}-\sigma_\mathrm{min})/(N-1)$. Note
that $\sigma_\mathrm{max}/\sigma_\mathrm{min} \to \infty$ at
$\delta_{\infty}\!=\!1/\sqrt{3}$.

The numerical investigation of equilibrium properties of fluids is
limited by the practical impossibility to equilibrate when either the
relaxation time $\tau$ within the fluid phase, or the freezing time
$t_\mathrm{fr}$~\footnote{We define $t_\mathrm{fr}$ as the
characteristic time for the first tunneling from the liquid to the
solid phase.  } become comparable with the time scale of the
simulation. In order to achieve equilibrium we straightforwardly adapt
to model~(\ref{potential}) the local swap MC algorithm
~\cite{Fernandez06,Fernandez06c}, which
outperforms~\cite{Fernandez06b,Fernandez06c} other algorithms, such as
the non-local swap MC~\cite{grigera01} and the density of states
MC~\cite{yan04}. In fact, the standard method of studying a
first-order transition in a system of finite size, $L$, looks for a
double peak in the histogram of the internal energy per particle $e$,
accompanied by a peak of the specific heat
$C_V$~\cite{Challa86,Lee90}. This procedure is extremely demanding on
the quality of the statistical sampling, and has been scarcely used
(if at all) for glass-forming liquids or colloids. Fortunately, the
local swap MC allows us to employ it.

\begin{figure}
\includegraphics[angle=0,width=\columnwidth,trim=10 20 20 9]{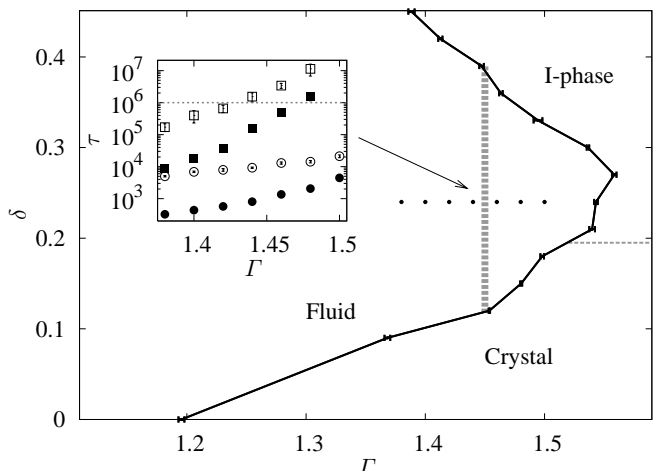}
\caption{The $\Gamma-\delta$ phase diagram shows three phases: a fluid
  at high temperatures, a crystal, and a inhomogeneous solid
  (I-phase). The boundary between the crystal and the I-phase lies at
  $0.18\!<\!\delta_\mathrm{c}\!<\!0.21$. The vertical line in the
  fluid phase is the kinetic glass transition. ({\em Inset})
  Integrated relaxation time~\cite{Fernandez06}, $\tau$, for $e$
  (full) and ${\cal F}$ (open), both for standard (squares) and local
  swap (circles) MC, for $N\!=\!256$. The kinetic glass transition
  arises when $\tau\! \sim\! 10^6$ {\em standard} MC
  steps.\protect{\label{PHASEDIAGRAM}}}
\vglue -3 mm
\end{figure}

\begin{figure}
\includegraphics[angle=0,width=\columnwidth,trim=15 15 21 9]{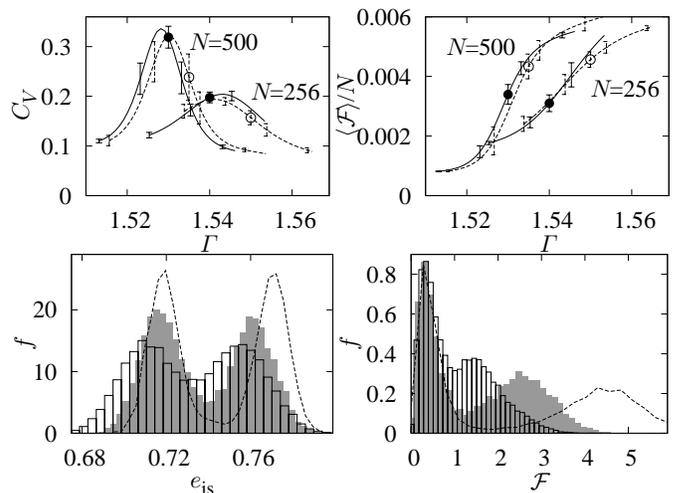} 
\caption{ Equilibrium quantities for $\delta\!=\!0.24$ ({\em
  top-left}) Specific heat vs. $\Gamma$, from two simulations, one at
  the size dependent critical point (full symbols denote the actual
  simulations, full lines coming from histogram reweighting),
  the other at a higher $\Gamma$ (open symbols, and dashed lines).
  The maximum of the specific heat scales with $N$. ({\em Top-right})
  $\langle{\cal F}\rangle/N$ vs. $\Gamma$ for the same simulations of
  top--left panel. ({\em Bottom-left}) Normalized histogram, $f$, of
  the energy of the inherent structures, for $N\!=\!256$ (empty) and
  $N\!=\!500$ (filled), at the $\Gamma$ value where the peaks of the
  instantaneous energy have equal
  height~\protect{\cite{Lee90}}. Inherent structures were found every
  $10^5$ MC steps.  ({\em Bottom-right}) As Bottom-left, for ${\cal
  F}$.  In the fluid phase ${\cal F}$ is ${\cal O}(1)$ (in the solid,
  ${\cal F}$ is ${\cal O}(N)$). We also show (dashed line)
  $e_\mathrm{is}$ and ${\cal F}$ histograms for $N=864$, where only
  two back and forth tunneling events between the liquid and solid
  phase were observed. Yet, data nicely agree with the predicted
  $N$-scaling.  \protect{\label{EQUILIBRIO}}}
\vglue -5 mm
\end{figure}

The thermalization issue needs to be addressed in three different
regimes, see Fig.~\ref{PHASEDIAGRAM}: the liquid phase, at the phase
coexistence line, and the solid phase. The swap algorithm avoids the
cage effect, thus equilibrating easily the whole liquid phase (it
reduces $\tau$ by two orders of magnitude as compared with standard
MC, see Fig.~\ref{PHASEDIAGRAM}--inset). Thermalization in the deep
solid phase, not attempted in this work, would require a different
approach. At the phase coexistence line, our criterion for
thermalization required the observation of tenths of back and forth
tunneling events between the liquid and solid phase. The difficulty
for meeting the criterion grows dramatically with the number of
particles ($\sim\exp[\Sigma N^{2/3}]$, $\Sigma$ being the liquid-solid
surface tension). A stronger, more quantitative check is the
consistency of the histogram reweighting~\footnote{The average for a
$N$ particles system at temperature $1/(\beta+\delta\beta)$, of any
function $A[\{x_i\}]$, can be obtained~\cite{Falcioni82,Ferrenberg89}
from mean-values at temperature $1/\beta$ through the identity
$\langle A\rangle_{\beta+\delta\beta} \!=\! \langle
A\exp[-N\delta\beta e]\rangle_{\beta}/\langle \exp[-N\delta\beta
e]\rangle_{\beta} \,,$ ($e$ is the energy per particle), that, in the
$\delta\beta\to 0$ limit, yields the (differential)
Fluctuation-Dissipation Theorem $\partial_\beta\langle
A\rangle_{\beta}\!=\!N[\langle A\,e\rangle_\beta- \langle
A\rangle_\beta\langle e\rangle_\beta]\,.$ Thus, the reweighting method
exploits an (integral) Fluctuation-Dissipation Theorem, and
consistency in the extrapolation (in practice, $\delta\beta\lesssim
[NC_V]^{-1/2}$) is a strong thermalization test.\label{EXPLICACIONRW}}.
We equilibrated $N\!=\!256$ particles for $\delta\geq 0.1$ (and for
the monodisperse system) and $N\!=\!500$ for $\delta\geq
0.21$. However, for local swap MC $t_\mathrm{fr}$ remains between
$10^6$ and $10^7$ MC steps for all $\delta$, while for standard MC it
grows beyond $10^9$ steps for $\delta>0.12$.

To detect the possible existence of large density fluctuations we focus
on ${\cal S}(\bm q) \equiv \frac{1}{N}\sum_{i,j} \mathrm{exp}[{\mathrm
i}\bm q\cdot (\bm r_i-\bm r_j)]$ ($\bm r_i$ is the position of the
$i$-th particle). Note that $\langle {\cal S}(\bm{q})\rangle$ is the
static structure factor. The longest wavelengths fluctuations are
studied through ${\cal F}\equiv [{\cal S}(\frac{2\pi}{L},0,0)+ {\cal
S}(0,\frac{2\pi}{L},0)+{\cal S}(0,0,\frac{2\pi}{L})]/3\,$. In fact, in
an homogeneous liquid or crystal phase we expect ${\cal F}$ to be of
order $1$, but for a macroscopically inhomogeneous phase it becomes of
order $N$.

In Fig.~\ref{EQUILIBRIO} we show as an example the results obtained
for $\delta\!=\!0.24$ where an evidence for a freezing transition is
presented. The specific-heat displays a peak of height proportional to
$N$ while, as usual for small systems, its position suffers a strong
finite size shift. We found convenient to use the histogram of the
energy per particle of the inherent structures $e_\mathrm{IS}$,
i.e. the minima of the potential energy~\cite{Stillinger83}. The
advantages are twofold (Fig.~\ref{EQUILIBRIO}): it largely absorbs the
effects of the finite-system shift of the critical temperature (so
that the position of the peaks is almost $N$ independent) and it makes
each peak narrower.

Note that $\delta\!=\!0.24$ is much higher than the terminal
polydispersity $\delta_\mathrm{t}$ reported in experiments and
simulations~\cite{Pusey86,Santen01}. Actually, a freezing transition
arises in the full range of $\delta$ studied. The line of
phase-coexistence, as located by the arising of a double peak for
$N\!=\!256$, between the fluid and the solid phase is shown in
Fig.~\ref{PHASEDIAGRAM}.

For $\delta\!=\!0$, we find a body-centered cubic (bcc) crystal, as
expected for modest $N$~\cite{WOLDE95}. Indeed, ${\cal S}(\bm{q})$
displays peaks of order $N$ at $\bm{q} \sim \frac{2\pi}{a}(1,1,0),
\frac{2\pi}{a}(0,1,1), \frac{2\pi}{a}(1,0,1)$ with lattice spacing $a
\sim 0.78$. The same Bragg peaks are found at $\delta\!=\!0.12$,
broadened due to disorder~\footnote{See~\cite{Megen06} for recent
experiments at low polydispersity.} (as compared with $\delta\!=\!0$,
the intensity reduces by a factor 1/4, but it still increases linearly
with $N$).

\begin{figure}
\includegraphics[angle=0,width=\columnwidth,trim=12 15 18 0]{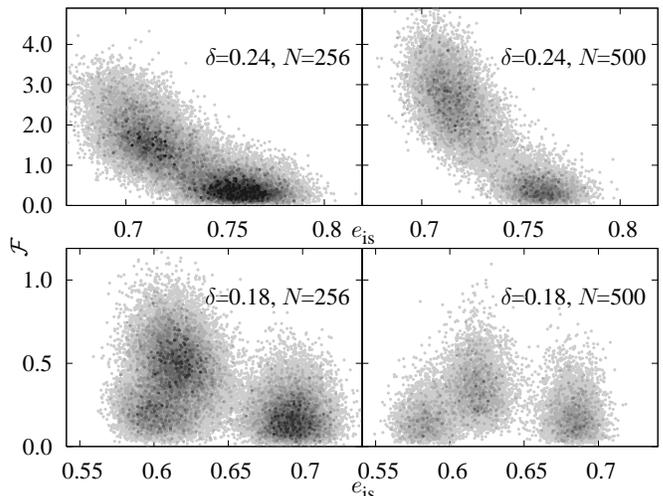}
\caption{Scatter plot of ${\cal F}$ vs. $e_\mathrm{IS}$ below (bottom)
and above (top) the I-phase-crystal transition line at the
liquid-solid phase coexistence, for $N\!=\!256$ (left) and $N\!=\!500$
(right). The number of points is $\sim 45000$ ($N\!=\!256$) or $\sim
15000$ ($N\!=\!500$). The darker the shade of gray, the higher the
density of points.\protect{\label{FUERA-EQUILIBRIO}}}
\end{figure}

\begin{figure}
\includegraphics[angle=0,width=\columnwidth,trim=10 15 5 10]{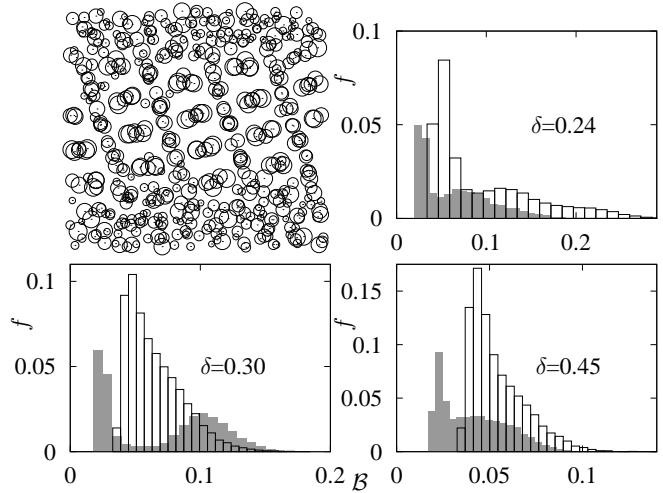}
\caption{ (Top-left) A $\delta\!=\!0.3$ solid configuration of 500
particles, at phase coexistence, as projected into the XY plane
(circles are centered at particles positions, their diameter being
proportional to particle sizes). Larger particles form the crystalline
central band
(the Bragg peaks's analysis suggests a FCC ordering). 
Normalized histograms of ${\cal B}$ for $\delta\!=\!0.24,0.3,0.45$
with 256 (empty) and 500 (filled) particles.\protect{\label{BRAGG}}}
\vglue -3 mm
\end{figure}

Interestingly, for $\delta\!=\!0.24$ the histogram of ${\cal F}$
(Fig.~\ref{EQUILIBRIO}) develops a double peak at the freezing
point. Left-peak's position is $N$-independent, as expected for the
liquid phase (see also the scatter plot in
Fig.~\ref{FUERA-EQUILIBRIO}). On the other hand, the second peak
(corresponding to the solid) shifts right proportionally to $N$,
revealing that the solid phase is not a standard crystalline one. For
$N\!=\!500$ and $\delta \lesssim 0.2$ it is risky to use histogram
techniques, as the tunneling from solid to liquid is harder to observe
than the reverse liquid to solid tunneling, thus we show in
Fig.~\ref{FUERA-EQUILIBRIO} a scatter plot.  It is clear that in the
crystals we found at $\delta\!=\!0.18$ (corresponding to the lower
values of $e_\mathrm{IS}$ in Fig.~\ref{FUERA-EQUILIBRIO}) ${\cal F}$
takes a $N$-independent value~\footnote{The $e_\mathrm{IS}$ dispersion
at $\delta\!=\!0.18$ crystals in Fig.~\ref{FUERA-EQUILIBRIO} is due
to the interplay between spatial orientation and 
periodic boundary conditions.}.  
Summarizing, at high polydispersities we have found a
freezing transition from the liquid to a solid inhomogeneous phase,
while at low polydispersities the low temperature high density phase
(large $\Gamma$'s) is a standard homogeneous crystal. The study of the
transition between the crystal and the I-phase is left for future
work. To gain some insight about the I-phase we measured the intensity
of the Bragg peaks for the inherent structures in the solid
phase. More precisely, we define ${\cal B}$ as the
maximum~\footnote{Maximum over $\bm q\!=\!\frac{2\pi}{L}(n_1,n_2,n_3),
|n_i|\leq 20$ and $\bm q\neq 0$.\label{EXPLICACION}} of ${\cal S} (\bm
q)$. In a crystal ${\cal B}$ is of order $N$ (the corresponding $\bm
q$ is in the reciprocal lattice), while in a fluid ${\cal B}$ is
always of order $1$. In Fig.~\ref{BRAGG} we show the histogram of
${\cal B}$ along the phase coexistence line both for $N\!=\!256$ and
$N\!=\!500$.  For every $\delta$ we find a double peak structure. The
solid's peak shifts right proportionally $N$. Thus $\langle {\cal
B}\rangle$ and $\langle {\cal F}\rangle$ provide a classification of
the $\Gamma-\delta$ plane: $\langle {\cal B}\rangle$ distinguishes the
solid from the fluid phase, while $\langle {\cal F}\rangle$ is of
order $N$ only in the I-phase.

The solid peak in Fig.~\ref{BRAGG} shifts left with increasing
$\delta$ (at $\delta\!=\!0.45$ the two peaks are hardly resolved). The
I-phase might signal either fractionation~\cite{Bartlett98,Kofke99} or
phase coexistence of a crystal with an amorphous solid
(Fig.~\ref{BRAGG}, top-left). Further work is needed to clarify this
point.

The existence of a freezing transition for all $\delta$ rules out the
terminal polydispersity scenario.  But even if a stable solid phase
exists for large $\delta$, it might be {\em dynamically} inaccessible
on experimental time-scales. In order to detect the occurrence of the
kinetic glass transition on the $\delta-\Gamma$ plane we adopted
standard MC simulations which (at variance with local swap MC), mimic
the real dynamics of the system by modeling it as a Brownian motion.
Note that the correspondence between the single step of standard MC
and the physical time units widely differ for liquids and colloids,
being $\sim 10^{-13}$ secs. in the former case and $\sim 10^{-2}$
secs. in the latter one~\cite{simeonova04}.

We have located the kinetic glass transition in
Fig.~\ref{PHASEDIAGRAM} as the point where $\tau$ reaches $10^6$
(standard) MC steps. In the case of colloids, this roughly corresponds
to a relaxation time of three hours, hence our kinetic glass
transition corresponds to the experimental one. On the other hand,
since in the case of liquids $10^6$ MC steps $\sim 10^{-7}$ secs., our
kinetic glass transition is rather close to the mode-coupling
transition~\cite{Goetze92}. Indeed, for most molecular and polymeric
glass-formers $\tau$ at the mode-coupling transition lies between
$10^{-7.5}$ and $10^{-6.5}$ secs.~\cite{Novikov03}. By the way, since
$\tau$ grows by many order of magnitudes in a narrow range, for
liquids the difference in $\Gamma$ between the experimental and the
mode coupling transition will not be larger than
$5\%$~\cite{Rossler94}. In the inset of Fig.~\ref{PHASEDIAGRAM} we
observe that the kinetic glass transition is at
$\Gamma_\mathrm{c}\simeq 1.45$ for $\delta\!=\!0.24$. This value coincides
with the one found for the soft-spheres binary mixtures (i.e. a
bimodal $P(\sigma)$) with $\delta\!=\!0.09$~\cite{Bernu87} and
$\delta\!=\!0.16$~\cite{Yu04}. This suggests that $\Gamma_\mathrm{c}$ is
independent on $\delta$. The vertical line at $\Gamma_\mathrm{c}\!=\!1.45$
intersects the freezing transition line at two {\em intersection
points}, creating a sort of no man's land where equilibrium is
experimentally unreachable on human time-scales. In fact, the freezing
time $t_\mathrm{fr}$ depends on $\delta$, growing tremendously along
the phase coexistence line.  When approaching the first intersection
point ($\delta \approx 0.12$, see Fig.~\ref{PHASEDIAGRAM}), for
standard MC we only know that $t_\mathrm{fr}$ becomes larger than
$10^9$ steps. For $\delta > 0.12$ our standard MC simulations did not
find the solid phase. We thus argue that the tremendous slowing down
both of $\tau$ and $t_\mathrm{fr}$ makes the solid phase for $\delta
\gtrsim 0.12$ experimentally unattainable.

We studied the equilibrium phase diagram in the $\delta-\Gamma$ plane
for a polydisperse soft sphere model with emphasis on the first order
freezing transition. Optimized MC algorithms and Finite Size Scaling
analysis were crucial to obtain equilibrium properties.  We found two
different solid phases, a standard crystal at small $\delta$ and a
highly inhomogeneous phase at large $\delta$.  Moreover, for $\delta$
lying between $0.12$ and $0.38$ the glass transition is no longer
preempted by the freezing transition, making this range very promising
for the theoretical study of glasses. In experiments non-equilibrium
effects should dominate the whole region $\delta \gtrsim 0.12$, where
either the kinetic glass transitions or the growth of the freezing
time-scale are expected to hide the freezing transition. This is in
quantitative agreement with the findings for colloids~\cite{Pusey86}
suggesting that the polydisperse soft-spheres model~(\ref{potential})
could catch the features both of molecular glass-formers and of
colloids.

We thank V. Erokhin and F. Zamponi for discussions.  
We were supported by MEC (Spain), through contracts BFM2003-08532,
FIS2004-05073, FPA2004-02602 and by BSCH---UCM. The CPU time utilized
(at BIFI and CINECA) amounts to 10 years of 3 GHz PentiumIV.


\bibliography{biblio}

\begin{thebibliography}{29}
\expandafter\ifx\csname natexlab\endcsname\relax\def\natexlab#1{#1}\fi
\expandafter\ifx\csname bibnamefont\endcsname\relax
  \def\bibnamefont#1{#1}\fi
\expandafter\ifx\csname bibfnamefont\endcsname\relax
  \def\bibfnamefont#1{#1}\fi
\expandafter\ifx\csname citenamefont\endcsname\relax
  \def\citenamefont#1{#1}\fi
\expandafter\ifx\csname url\endcsname\relax
  \def\url#1{\texttt{#1}}\fi
\expandafter\ifx\csname urlprefix\endcsname\relax\def\urlprefix{URL }\fi
\providecommand{\bibinfo}[2]{#2}
\providecommand{\eprint}[2][]{\url{#2}}

\bibitem[{\citenamefont{Sciortino and Tartaglia}(2005)}]{Sciortino05}
\bibinfo{author}{\bibfnamefont{F.}~\bibnamefont{Sciortino}} \bibnamefont{and}
  \bibinfo{author}{\bibfnamefont{P.}~\bibnamefont{Tartaglia}},
  \bibinfo{journal}{Adv. Phys.} \textbf{\bibinfo{volume}{54}},
  \bibinfo{pages}{471} (\bibinfo{year}{2005}).

\bibitem[{\citenamefont{Fasolo and Sollich}(2004)}]{Fasolo04}
\bibinfo{author}{\bibfnamefont{M.}~\bibnamefont{Fasolo}} \bibnamefont{and}
  \bibinfo{author}{\bibfnamefont{P.}~\bibnamefont{Sollich}},
  \bibinfo{journal}{Phys. Rev. E} \textbf{\bibinfo{volume}{70}},
  \bibinfo{pages}{041410} (\bibinfo{year}{2004}).

\bibitem[{\citenamefont{Chaudhuri et~al.}(2005)\citenamefont{Chaudhuri,
  Karmakar, Dasgupta, Krishnamurthy, and Sood}}]{Chaudhuri05}
\bibinfo{author}{\bibfnamefont{P.}~\bibnamefont{Chaudhuri}},
  \bibinfo{author}{\bibfnamefont{S.}~\bibnamefont{Karmakar}},
  \bibinfo{author}{\bibfnamefont{C.}~\bibnamefont{Dasgupta}},
  \bibinfo{author}{\bibfnamefont{H.~R.} \bibnamefont{Krishnamurthy}},
  \bibnamefont{and} \bibinfo{author}{\bibfnamefont{A.~K.} \bibnamefont{Sood}},
  \bibinfo{journal}{Phys. Rev. Lett.} \textbf{\bibinfo{volume}{95}},
  \bibinfo{eid}{248301} (pages~\bibinfo{numpages}{4}) (\bibinfo{year}{2005}).

\bibitem[{\citenamefont{Bartlett}(1998)}]{Bartlett98}
\bibinfo{author}{\bibfnamefont{P.}~\bibnamefont{Bartlett}},
  \bibinfo{journal}{J. Chem. Phys.} \textbf{\bibinfo{volume}{109}},
  \bibinfo{pages}{10970} (\bibinfo{year}{1998}).

\bibitem[{\citenamefont{Kofke and Bolhuis}(1999)}]{Kofke99}
\bibinfo{author}{\bibfnamefont{D.~A.} \bibnamefont{Kofke}} \bibnamefont{and}
  \bibinfo{author}{\bibfnamefont{P.~G.} \bibnamefont{Bolhuis}},
  \bibinfo{journal}{Phys. Rev. E} \textbf{\bibinfo{volume}{59}},
  \bibinfo{pages}{618} (\bibinfo{year}{1999}).

\bibitem[{\citenamefont{Auer and Frenkel}(2001)}]{Auer01}
\bibinfo{author}{\bibfnamefont{S.}~\bibnamefont{Auer}} \bibnamefont{and}
  \bibinfo{author}{\bibfnamefont{D.}~\bibnamefont{Frenkel}},
  \bibinfo{journal}{Nature} \textbf{\bibinfo{volume}{413}},
  \bibinfo{pages}{711} (\bibinfo{year}{2001}).

\bibitem[{\citenamefont{Pusey and van Megen}(1986)}]{Pusey86}
\bibinfo{author}{\bibfnamefont{P.~N.} \bibnamefont{Pusey}} \bibnamefont{and}
  \bibinfo{author}{\bibfnamefont{W.}~\bibnamefont{van Megen}},
  \bibinfo{journal}{Nature} \textbf{\bibinfo{volume}{320}},
  \bibinfo{pages}{340} (\bibinfo{year}{1986}).

\bibitem[{\citenamefont{Dullens and Kegel}(2004)}]{Dullens04}
\bibinfo{author}{\bibfnamefont{R.~P.~A.} \bibnamefont{Dullens}}
  \bibnamefont{and} \bibinfo{author}{\bibfnamefont{W.~K.} \bibnamefont{Kegel}},
  \bibinfo{journal}{Phys. Rev. Lett.} \textbf{\bibinfo{volume}{92}},
  \bibinfo{eid}{195702} (pages~\bibinfo{numpages}{4}) (\bibinfo{year}{2004}).

\bibitem[{\citenamefont{Debenedetti}(1997)}]{DeBenedetti97}
\bibinfo{author}{\bibfnamefont{P.~G.} \bibnamefont{Debenedetti}},
  \emph{\bibinfo{title}{Metastable Liquids}} (\bibinfo{publisher}{Princeton
  University Press}, \bibinfo{year}{1997}).

\bibitem[{\citenamefont{Hansen and McDonald}(1986)}]{hansen}
\bibinfo{author}{\bibfnamefont{J.~P.} \bibnamefont{Hansen}} \bibnamefont{and}
  \bibinfo{author}{\bibfnamefont{I.~R.} \bibnamefont{McDonald}},
  \emph{\bibinfo{title}{Theory of Simple Liquids}}
  (\bibinfo{publisher}{Academic Press}, \bibinfo{address}{San Diego},
  \bibinfo{year}{1986}).

\bibitem[{\citenamefont{Santen and Krauth}(2001)}]{Santen01}
\bibinfo{author}{\bibfnamefont{L.}~\bibnamefont{Santen}} \bibnamefont{and}
  \bibinfo{author}{\bibfnamefont{W.}~\bibnamefont{Krauth}},
  \bibinfo{howpublished}{condmat/0107459} (\bibinfo{year}{2001}).

\bibitem[{\citenamefont{Fern\'andez
  et~al.}(2006{\natexlab{a}})\citenamefont{Fern\'andez, Mart\'in-Mayor, and
  Verrocchio}}]{Fernandez06}
\bibinfo{author}{\bibfnamefont{L.~A.} \bibnamefont{Fern\'andez}},
  \bibinfo{author}{\bibfnamefont{V.}~\bibnamefont{Mart\'in-Mayor}},
  \bibnamefont{and}
  \bibinfo{author}{\bibfnamefont{P.}~\bibnamefont{Verrocchio}},
  \bibinfo{journal}{Phys. Rev. E} \textbf{\bibinfo{volume}{73}},
  \bibinfo{pages}{020501(R)} (\bibinfo{year}{2006}{\natexlab{a}}).

\bibitem[{\citenamefont{Fern\'andez
  et~al.}(2006{\natexlab{b}})\citenamefont{Fern\'andez, Mart\'in-Mayor, and
  Verrocchio}}]{Fernandez06c}
\bibinfo{author}{\bibfnamefont{L.~A.} \bibnamefont{Fern\'andez}},
  \bibinfo{author}{\bibfnamefont{V.}~\bibnamefont{Mart\'in-Mayor}},
  \bibnamefont{and}
  \bibinfo{author}{\bibfnamefont{P.}~\bibnamefont{Verrocchio}},
  \bibinfo{journal}{Philosophical Magazine} \textbf{\bibinfo{volume}{87}},
  \bibinfo{pages}{581} (\bibinfo{year}{2006}{\natexlab{b}}).

\bibitem[{\citenamefont{Fern\'andez et~al.}(2005)\citenamefont{Fern\'andez,
  Mart\'in-Mayor, and Verrocchio}}]{Fernandez06b}
\bibinfo{author}{\bibfnamefont{L.~A.} \bibnamefont{Fern\'andez}},
  \bibinfo{author}{\bibfnamefont{V.}~\bibnamefont{Mart\'in-Mayor}},
  \bibnamefont{and}
  \bibinfo{author}{\bibfnamefont{P.}~\bibnamefont{Verrocchio}}
  (\bibinfo{publisher}{AIP}, \bibinfo{address}{Melville, New York},
  \bibinfo{year}{2005}), no. \bibinfo{number}{832} in \bibinfo{series}{AIP
  Conference Proceedings Series}, pp. \bibinfo{pages}{128--133}.

\bibitem[{\citenamefont{Grigera and Parisi}(2001)}]{grigera01}
\bibinfo{author}{\bibfnamefont{T.~S.} \bibnamefont{Grigera}} \bibnamefont{and}
  \bibinfo{author}{\bibfnamefont{G.}~\bibnamefont{Parisi}},
  \bibinfo{journal}{Phys. Rev. E} \textbf{\bibinfo{volume}{63}},
  \bibinfo{pages}{045102(R)} (\bibinfo{year}{2001}).

\bibitem[{\citenamefont{Yan et~al.}(2004)\citenamefont{Yan, Jain, and
  de~Pablo}}]{yan04}
\bibinfo{author}{\bibfnamefont{Q.}~\bibnamefont{Yan}},
  \bibinfo{author}{\bibfnamefont{T.~S.} \bibnamefont{Jain}}, \bibnamefont{and}
  \bibinfo{author}{\bibfnamefont{J.~J.} \bibnamefont{de~Pablo}},
  \bibinfo{journal}{Phys. Rev. Lett.} \textbf{\bibinfo{volume}{92}},
  \bibinfo{eid}{235701} (pages~\bibinfo{numpages}{4}) (\bibinfo{year}{2004}).

\bibitem[{\citenamefont{Challa et~al.}(1986)\citenamefont{Challa, Landau, and
  Binder}}]{Challa86}
\bibinfo{author}{\bibfnamefont{M.~S.~S.} \bibnamefont{Challa}},
  \bibinfo{author}{\bibfnamefont{D.~P.} \bibnamefont{Landau}},
  \bibnamefont{and} \bibinfo{author}{\bibfnamefont{K.}~\bibnamefont{Binder}},
  \bibinfo{journal}{Phys. Rev. B} \textbf{\bibinfo{volume}{34}},
  \bibinfo{pages}{1841} (\bibinfo{year}{1986}).

\bibitem[{\citenamefont{Lee and Kosterlitz}(1990)}]{Lee90}
\bibinfo{author}{\bibfnamefont{J.}~\bibnamefont{Lee}} \bibnamefont{and}
  \bibinfo{author}{\bibfnamefont{J.~M.} \bibnamefont{Kosterlitz}},
  \bibinfo{journal}{Phys. Rev. Lett.} \textbf{\bibinfo{volume}{65}},
  \bibinfo{pages}{137} (\bibinfo{year}{1990}).

\bibitem[{\citenamefont{Stillinger and Weber}(1983)}]{Stillinger83}
\bibinfo{author}{\bibfnamefont{F.~H.} \bibnamefont{Stillinger}}
  \bibnamefont{and} \bibinfo{author}{\bibfnamefont{T.~A.} \bibnamefont{Weber}},
  \bibinfo{journal}{Phys. Rev. A} \textbf{\bibinfo{volume}{28}},
  \bibinfo{pages}{2408} (\bibinfo{year}{1983}).

\bibitem[{\citenamefont{ten Wolde et~al.}(1995)\citenamefont{ten Wolde,
  Ruiz-Montero, and Frenkel}}]{WOLDE95}
\bibinfo{author}{\bibfnamefont{P.~R.} \bibnamefont{ten Wolde}},
  \bibinfo{author}{\bibfnamefont{M.~J.} \bibnamefont{Ruiz-Montero}},
  \bibnamefont{and} \bibinfo{author}{\bibfnamefont{D.}~\bibnamefont{Frenkel}},
  \bibinfo{journal}{Phys. Rev. Lett.} \textbf{\bibinfo{volume}{75}},
  \bibinfo{pages}{2714} (\bibinfo{year}{1995}).

\bibitem[{\citenamefont{Simeonova and Kegel}(2004)}]{simeonova04}
\bibinfo{author}{\bibfnamefont{N.~B.} \bibnamefont{Simeonova}}
  \bibnamefont{and} \bibinfo{author}{\bibfnamefont{W.~K.} \bibnamefont{Kegel}},
  \bibinfo{journal}{Phys. Rev. Lett.} \textbf{\bibinfo{volume}{93}},
  \bibinfo{eid}{035701} (pages~\bibinfo{numpages}{4}) (\bibinfo{year}{2004}).

\bibitem[{\citenamefont{G{\"o}tze and Sj{\"o}gren}(1992)}]{Goetze92}
\bibinfo{author}{\bibfnamefont{W.}~\bibnamefont{G{\"o}tze}} \bibnamefont{and}
  \bibinfo{author}{\bibfnamefont{L.}~\bibnamefont{Sj{\"o}gren}},
  \bibinfo{journal}{Rep. Prog. Phys.} \textbf{\bibinfo{volume}{55}},
  \bibinfo{pages}{241} (\bibinfo{year}{1992}).

\bibitem[{\citenamefont{Novikov and Sokolov}(2003)}]{Novikov03}
\bibinfo{author}{\bibfnamefont{V.~N.} \bibnamefont{Novikov}} \bibnamefont{and}
  \bibinfo{author}{\bibfnamefont{A.~P.} \bibnamefont{Sokolov}},
  \bibinfo{journal}{Phys. Rev. E} \textbf{\bibinfo{volume}{67}},
  \bibinfo{eid}{031507} (pages~\bibinfo{numpages}{6}) (\bibinfo{year}{2003}).

\bibitem[{\citenamefont{R\"ossler et~al.}(1994)\citenamefont{R\"ossler,
  Sokolov, Kisliuk, and Quitmann}}]{Rossler94}
\bibinfo{author}{\bibfnamefont{E.}~\bibnamefont{R\"ossler}},
  \bibinfo{author}{\bibfnamefont{A.~P.} \bibnamefont{Sokolov}},
  \bibinfo{author}{\bibfnamefont{A.}~\bibnamefont{Kisliuk}}, \bibnamefont{and}
  \bibinfo{author}{\bibfnamefont{D.}~\bibnamefont{Quitmann}},
  \bibinfo{journal}{Phys. Rev. B} \textbf{\bibinfo{volume}{49}},
  \bibinfo{pages}{14967} (\bibinfo{year}{1994}).

\bibitem[{\citenamefont{Bernu et~al.}(1987)\citenamefont{Bernu, Hansen,
  Hiwatari, and Pastore}}]{Bernu87}
\bibinfo{author}{\bibfnamefont{B.}~\bibnamefont{Bernu}},
  \bibinfo{author}{\bibfnamefont{J.~P.} \bibnamefont{Hansen}},
  \bibinfo{author}{\bibfnamefont{Y.}~\bibnamefont{Hiwatari}}, \bibnamefont{and}
  \bibinfo{author}{\bibfnamefont{G.}~\bibnamefont{Pastore}},
  \bibinfo{journal}{Phys. Rev. A} \textbf{\bibinfo{volume}{36}},
  \bibinfo{pages}{4891} (\bibinfo{year}{1987}).

\bibitem[{\citenamefont{Yu and Carruzzo}(2004)}]{Yu04}
\bibinfo{author}{\bibfnamefont{C.~C.} \bibnamefont{Yu}} \bibnamefont{and}
  \bibinfo{author}{\bibfnamefont{H.~M.} \bibnamefont{Carruzzo}},
  \bibinfo{journal}{Phys. Rev. E} \textbf{\bibinfo{volume}{69}},
  \bibinfo{pages}{051201} (\bibinfo{year}{2004}).

\bibitem[{\citenamefont{Falcioni et~al.}(1982)\citenamefont{Falcioni, Marinari,
  Paciello, Parisi, and Taglienti}}]{Falcioni82}
\bibinfo{author}{\bibfnamefont{M.}~\bibnamefont{Falcioni}},
  \bibinfo{author}{\bibfnamefont{E.}~\bibnamefont{Marinari}},
  \bibinfo{author}{\bibfnamefont{M.~L.} \bibnamefont{Paciello}},
  \bibinfo{author}{\bibfnamefont{G.}~\bibnamefont{Parisi}}, \bibnamefont{and}
  \bibinfo{author}{\bibfnamefont{B.}~\bibnamefont{Taglienti}},
  \bibinfo{journal}{Phys. Lett. B} \textbf{\bibinfo{volume}{108}},
  \bibinfo{pages}{331} (\bibinfo{year}{1982}).

\bibitem[{\citenamefont{Ferrenberg and Swendsen}(1989)}]{Ferrenberg89}
\bibinfo{author}{\bibfnamefont{A.~M.} \bibnamefont{Ferrenberg}}
  \bibnamefont{and} \bibinfo{author}{\bibfnamefont{R.~H.}
  \bibnamefont{Swendsen}}, \bibinfo{journal}{Phys. Rev. Lett.}
  \textbf{\bibinfo{volume}{63}}, \bibinfo{pages}{1195} (\bibinfo{year}{1989}).

\bibitem[{\citenamefont{Sch\"ope et~al.}(2006)\citenamefont{Sch\"ope, Bryant,
  and van Megen}}]{Megen06}
\bibinfo{author}{\bibfnamefont{H.~J.} \bibnamefont{Sch\"ope}},
  \bibinfo{author}{\bibfnamefont{G.}~\bibnamefont{Bryant}}, \bibnamefont{and}
  \bibinfo{author}{\bibfnamefont{W.}~\bibnamefont{van Megen}},
  \bibinfo{journal}{Phys. Rev. Lett.} \textbf{\bibinfo{volume}{96}},
  \bibinfo{eid}{175701} (pages~\bibinfo{numpages}{4}) (\bibinfo{year}{2006}).

\end{thebibliography}

\end{document}